\documentclass[10pt,conference]{IEEEtran}
\usepackage{cite}
\usepackage{amsmath,amssymb,amsfonts}
\usepackage{algorithmic}
\usepackage{graphicx}
\usepackage{textcomp}
\usepackage{color,xcolor}
\usepackage{colortbl}
\usepackage[hyphens]{url}
\usepackage{fancyhdr}
\usepackage[bookmarks=true,breaklinks=true,letterpaper=true,colorlinks,citecolor=blue,linkcolor=blue,urlcolor=blue]{hyperref}

\newcommand{\hpcayear}{2025}

\newcommand{\hpcasubmissionnumber}{NaN}
\title{EFFACT: A Highly Efficient Full-Stack FHE Acceleration Platform } 

\def\hpcacameraready{} 

\newcommand\hpcaauthors{Yi Huang*, Xinsheng Gong*, Xiangyu Kong, Dibei Chen, Jianfeng Zhu$\dagger$, Wenping Zhu, Liangwei Li,\\ Mingyu Gao, Shaojun Wei, Aoyang Zhang and Leibo Liu$\dagger$}
\newcommand\hpcaaffiliation{Tsinghua University\\ \ \
Equal Contribution*
\\ \ 
\ \ Corresponding Authors$\dagger$}
\newcommand\hpcaemail{\{yi-huang,jfzhu,liulb\}@tsinghua.edu.cn}

\author{
  \ifdefined\hpcacameraready
    \IEEEauthorblockN{\hpcaauthors{}}
      \IEEEauthorblockA{
        \hpcaaffiliation{} \\
        \hpcaemail{}
      }
  \else
    \IEEEauthorblockN{\normalsize{HPCA \hpcayear{} Submission
      \textbf{\#\hpcasubmissionnumber{}}} \\
      \IEEEauthorblockA{
        Confidential Draft \\
        Do NOT Distribute!!
      }
    }
  \fi 
}

\fancypagestyle{camerareadyfirstpage}{%
  \fancyhead{}
  
  \fancyhead[C]{
    \ifdefined\aeopen
    \parbox[][12mm][t]{13.5cm}{\hpcayear{} IEEE International Symposium on High-Performance Computer Architecture (HPCA)}    
    \else
      \ifdefined\aereviewed
      \parbox[][12mm][t]{13.5cm}{\hpcayear{} IEEE International Symposium on High-Performance Computer Architecture (HPCA)}
      \else
      \ifdefined\aereproduced
      \parbox[][12mm][t]{13.5cm}{\hpcayear{} IEEE International Symposium on High-Performance Computer Architecture (HPCA)}
      \else
      \parbox[][0mm][t]{13.5cm}{\hpcayear{} IEEE International Symposium on High-Performance Computer Architecture (HPCA)}
    \fi 
    \fi 
    \fi 
    \ifdefined\aeopen 
      \includegraphics[width=12mm,height=12mm]{ae-badges/open-research-objects.pdf}
    \fi 
    \ifdefined\aereviewed
      \includegraphics[width=12mm,height=12mm]{ae-badges/research-objects-reviewed.pdf}
    \fi 
    \ifdefined\aereproduced
      \includegraphics[width=12mm,height=12mm]{ae-badges/results-reproduced.pdf}
    \fi
  }
  \fancyfoot[C]{}
}
\begin{document}
\maketitle

\ifdefined\hpcacameraready 
  \thispagestyle{camerareadyfirstpage}
  \pagestyle{empty}
\else
  \thispagestyle{plain}
  \pagestyle{plain}
\fi

\newcommand{\hpcaheight}{0mm}
\ifdefined\eaopen
\renewcommand{\hpcaheight}{12mm}
\fi






\definecolor{yi}{RGB}{0,0,0}
\definecolor{gg}{RGB}{0,0,0}
\definecolor{gg_r}{RGB}{0,0,0}
\definecolor{revised}{RGB}{0,0,0}
\definecolor{rebuttal}{RGB}{0,0,0}


\begin{abstract}

Fully Homomorphic Encryption (FHE) is a set of powerful cryptographic schemes that allows computation to be performed directly on encrypted data with an unlimited depth. Despite FHE's promising in privacy-preserving computing, yet in most FHE schemes, ciphertext generally blows up thousands of times compared to the original message, and the massive amount of data load from off-chip memory for bootstrapping and privacy-preserving machine learning applications (such as HELR, ResNet-20), both degrade the performance of FHE-based computation. Several hardware designs have been proposed to address this issue, however, most of them require enormous resources and power. An acceleration platform with easy programmability, high efficiency, and low overhead is a prerequisite for practical application.

This paper proposes EFFACT, a highly efficient full-stack FHE acceleration platform with a compiler that provides comprehensive optimizations and vector-friendly hardware.
We start by examining the computational overhead across different real-world benchmarks to highlight the potential benefits of reallocating computing resources for efficiency enhancement. Then we make a design space exploration to find an optimal SRAM size with high utilization and low cost.
On the other hand, EFFACT features a novel optimization named streaming memory access which is proposed to enable high throughput with limited SRAMs.
Regarding the software-side optimization, we also propose a circuit-level function unit reuse scheme, to substantially reduce the computing resources without performance degradation.
Moreover, we design novel NTT and automorphism units that are suitable for a cost-sensitive and highly efficient architecture, leading to low area.
For generality, EFFACT is also equipped with an ISA and a compiler backend that can support several FHE schemes like CKKS, BGV, and BFV.

We provide both FPGA and ASIC versions of EFFACT. On account of our full stack design, FPGA-EFFACT outperforms the SOTA FPGA accelerators in gmean by 1.22$\times$. Meanwhile, ASIC-EFFACT shows increased improvements in terms of the performance per chip area and the performance per Watt compared with the SOTA ASIC works.

\end{abstract}

\section{Introduction}

Fully Homomorphic Encryption (FHE), due to its unique ability to compute encrypted data without requiring a secret key, is indispensable in the privacy-preserving computing field. 
In certain data privacy-sensitive scenarios like finance and medicine, FHE enables clients to access the computing and storage capabilities of cloud servers without disclosing confidential information.

{\color{gg_r}
FHE has been used in many applications to protect user's privacy data, e.g. ACGT base-pairs sequence in the genomic computing\cite{li_semi-parallel_2019}, the financial transaction\cite{han_logistic_2019}, and the image processing in CT\cite{cryptoeprint:2021/1688, georgieva_privacy-preserving_2019}. However, the amount of computations performed on FHE's encrypted data exceeds that of the unencrypted data. This is mainly because most FHE schemes are lattice-based (e.g., BGV, BFV, CKKS, TFHE)  \cite{brakerski_leveled_nodate,BFV,cheon_homomorphic_nodate,TFHE}, which utilizes the complexity of the Learning With Error (LWE) problem \cite{LWE} to ensure the security. Besides, FHE schemes necessitate extra operations for ciphertext maintenance to ensure the compactness and correctness of the scheme. These maintenance operations, such as key switching, rescale, and bootstrapping, are not efficient in CPU\cite{SOK_a}. Therefore, FHE applications usually need thousands of seconds to be completed on the CPU while the same applications on the unencrypted data only require milliseconds. To tackle this problem, various software methods have been explored, including optimizing FHE algorithms\cite{cid_full_2019, avanzi_full_2017, matsui_improved_2019} and developing highly optimized HE libraries for CPU or GPU-based systems\cite{cryptoeprint:2022/915, mouchet_lattigo_2020, halevi_design_nodate, chen_simple_nodate}. However, their performance enhancements remain inadequate to meet the demand of practical applications. In recent years, researchers have focused on creating Domain Specific Architecture (DSA) for FHE to make it more applicable to real-world scenarios.
}
Currently, efforts have been made to implement ASIC and FPGA designs for FHE. 
{\color{revised}The ASIC designs such as CraterLake \cite{samardzic_craterlake_2022}, BTS \cite{kim_bts_2022}, and ARK \cite{kim_ark_2022} provided more than 2 orders of magnitude speedup over GPU, showing a promising future for the large scale adoption of FHE. However, they have not explored the computation overhead and off-chip bandwidth efficiency. As a result, they require enormous resources, typically hundreds of megabytes of on-chip SRAM and tens of thousands of multipliers. {\color{revised}The huge on-chip resources incur large area consumption and huge energy consumption, which is expensive for commercial use and leads to low efficiency\cite{samardzic_craterlake_2022,sharp-kim}.}}
On the contrary, the FPGA solutions\cite{agrawal_fab_2022,yang_poseidon_nodate} are much more efficient due to their highly reused design. 
{\color{revised}However, the FPGA implementations only execute one HE operation every time due to the limited resources, and just a handful of scheduling or parallelism is explored, restricting their throughput.}
A practical design that features high throughput, flexibility, and high efficiency is demanded in the area of commercial FHE acceleration.

{\color{revised}MAD\cite{MAD} and SHARP\cite{sharp-kim} are the pioneers in designing a cost-sensitive and highly efficient architecture. 
SHARP gives an in-depth analysis of the word length to tackle the memory and NoC bottleneck, which is orthogonal to our proposal.
MAD proposes a novel caching scheme to explore data reuse and reduces the on-chip SRAM requirement by 16$\times$.}
However, MAD only looks into the on-chip SRAM optimization while still keeping the computing resources, its buffers and SRAM bandwidth as high as prior designs \cite{kim_ark_2022,samardzic_craterlake_2022,kim_bts_2022}, thus still requiring substantial area and energy consumption.
{\color{gg}Meanwhile, its caching scheme falls into hand-tuned data path scheduling within HE primitives and relies on the buffers on the computing resource side, revealing further optimizations.}
Therefore, there remains a significant design space for developing a cost-effective and highly efficient FHE accelerator.


{\color{revised}To alleviate those drawbacks, in this work, we first exhibit the possibility of re-distributing the computing resources by a detailed analysis of the proportion of different FHE operations at the residue polynomial level in real-world benchmarks. Then we also take a design exploration of the size of on-chip memory based on several trade-offs including performance, efficiency, and cost. Exploring computing resources and on-chip memory to achieve high efficiency and low cost while maintaining high throughput is an essential step in many accelerator designs\cite{dsagen,overgen,AHA,AURORA,Hansong,HASCO,soc,archgym}. However, such a resource-aware design is not well explored in prior FHE works.}

{\color{revised}Starting from the analysis, we build EFFACT, a highly \textbf{E}fficient \textbf{F}ull-stack \textbf{F}HE \textbf{AC}celeration pla\textbf{T}form to enhance the performance, area efficiency, and power efficiency of a cost-sensitive FHE accelerator. 
To obtain high speed with limited on-chip memory, we propose an automatic software-level optimization named streaming and its architecture-level support, in which the compiler finds the temporary data with less reuse and directly sends them from DRAM to function units without the buffering of SRAMs\cite{somogyi2006spatial,nowatzki2017stream}.
To enhance the area and power efficiency, we propose a circuit-level reuse scheme that judiciously uses both the NTT's multipliers and modular mult's multipliers to accelerate mult-accumulate (MAC) operations without performance degradation.
Meanwhile, we also devise specialized NTT and automorphism units to adapt to the cost-sensitive and highly efficient microarchitecture, further reducing computing resources.
}

{\color{revised}Without loss of generality, we also analyze the basic residue polynomial level operations in different FHE schemes including CKKS, BGV, and BFV, and design EFFACT's Instruction Set Architecture (ISA) and a compiler backend featuring automatic code optimization. 
Through the efforts of the ISA and compiler, our platform can flexibly support many kinds of FHE schemes and can be integrated into the state-of-the-art compiler frontends and analysis\cite{HEAANMLIR,static}.}

The contribution of this work can be summarized as follows:

\begin{itemize}
    \item We propose EFFACT, a full-stack FHE acceleration platform with generalized ISA and compiler backend, which runs near speed as resource-unconstrained designs with higher efficiency and low hardware overhead.
    
    \item We analyze the proportion of different FHE operations among different benchmarks and show the impact of different resource assignments on the efficiency and performance of the FHE accelerator.

    \item We propose a novel compiler optimization scheme on the software side with its architectural support that keeps high throughput while using extremely small on-chip memory.

    \item We devise a circuit-level reuse scheme for function units that reduces hardware redundancy without performance penalty and specialized resource-saving NTT and automorphism units. 
    
    \item We rigorously evaluate EFFACT in logistic regression and bootstrapping with both ASIC and FPGA versions. The experiments are performed by scaling the real FPGA runtime. Results are verified by comparing with Lattigo\cite{mouchet_lattigo_2020}. On account of our full stack design, FPGA-EFFACT outperforms the SOTA FPGA accelerators in gmean by 1.22$\times$. Meanwhile, ASIC-EFFACT shows increased improvements by $\geq$1.46$\times$ in terms of the performance per chip area and $\geq$1.48$\times$ in terms of the performance per Watt compared with the SOTA ASIC works.
\end{itemize}

\section{Background}
As mentioned above, there are different types of FHE schemes. In this section, we will take CKKS as an example to explain how FHE works. Relevant parameters and notations for the CKKS scheme can be found in Table~\ref{CKKS_param_table}. 
\begin{table}[h!]
  \centering
  \caption{CKKS Scheme Parameter and Notation}
  \vskip -2ex
  \small
  \scalebox{0.9}{
  \begin{tabular}{|l|l|}
    \hline
    \textbf{Notation} & \textbf{Description}\\
    \hline
    \hline
    \textbf{$N$} & Degree of cyclotomic ring\\
    \hline
    \textbf{$Q$} & Biggest modulus of ciphertext\\
    \hline
    \textbf{$P$} & Modulus product of all extension limbs\\
    \hline
    \textbf{$R_Q$} & Cyclotomic polynomial ring, $R_Q$ =$Z_Q[X]/(X^N+1)$\\
    \hline
    \textbf{$q_i$} & Modulus at level i of the modulus chain\\
    \hline
    \textbf{$L$} & Max level of ciphertext\\
    \hline
    \textbf{$l$} & Current level of ciphertext\\
    \hline
    \textbf{$dnum$} & Number of decompose digits\\
    \hline
    \textbf{$L_{boot}$} & Consumed level of bootstrapping \\
    \hline
    \textbf{$L_{CtS}$} & Consumed level of CtS in bootstrapping \\
    \hline
    \textbf{$L_{StC}$} & Consumed level of StC in bootstrapping \\
    \hline
    \textbf{$L_{EvalMod}$} & Consumed level of EvalMod in bootstrapping \\
    \hline
  \end{tabular}
  }
  \label{CKKS_param_table}
\end{table}
\subsection{RNS-CKKS FHE scheme}
CKKS, proposed by Cheon et al. \cite{cheon_homomorphic_nodate}, supports fixed-point real and complex data types and SIMD operation, which introduces approximate calculation into homomorphic encryption algorithm. It trades the loss of accuracy for a huge increase in computational efficiency compared to BGV/BFV schemes.

In EFFACT platform, we ignore the encoding/decoding and encrypting/decrypting steps of the scheme, since these are executed on the client side. Instead, we focus directly on the plaintext, ciphertext, and related homomorphic operations.
{\color{rebuttal} In CKKS, plaintext can be presented as a polynomial $m(X)=\sum_{i=0}^{N-1}m_iX^i$. The plaintext is an element of the cyclotomic polynomial ring $R_Q=Z_Q[X]/(X^N+1)$, which means all the N-degree polynomial plaintext's coefficients ($m_i$) fall within the range [0, $Q$-1] and the number of their coefficients is N. Typically, N is the power of 2. 
A single plaintext is packed (encoded) from a so-called message, which consists of a vector of N/2 complex numbers. Each \textbf{element} on the plaintext is called a \textbf{slot}.
The multiplication or addition of the message can be performed through the polynomial operations on the plaintext.
Then the plaintext ($m(X)\in R_Q$) will be encrypted into the ciphertext ($\textbf{ct}(X)\in R_{Q}^{2}$).
The ciphertext is represented as $\textbf{ct}(X) = (c_0(X), c_1(X))$ satisfying $c_0(X) = c_1(X)\cdot s(X)+\Delta m(X)+e(X)$, where $s(X)$ is the secret key, $c_1(X) \in R_Q$ is a random polynomial, $e(X) \in R_Q$ is a small error polynomial is added for security, and $\Delta$ is the scaling factor. The recovery of plaintext is conducted by $m'(X) = \textbf{ct}(X)\cdot (1,-s(X))=\Delta m(X)+e(X)$.}

RNS-CKKS scheme \cite{cid_full_2019, avanzi_full_2017, matsui_improved_2019} further improves the efficiency by using the Chinese Remainder Theorem (CRT) to decompose a big prime base (the Q base in the $R_Q$, which requires thousands of bits) into L small primes (denoted as $Q=\prod_{i=0}^{L-1}q_i$) with shorter bit width. {\color{rebuttal}Therefore, a polynomial $m(X)$ in the $R_Q$ can be represented using L \textbf{residue polynomials} or \textbf{limbs} in $R_{q_i}$ as $\{m_0[X]\in R_{q_0}, m_1[X]\in R_{q_1}, ..., m_{L-1}[X]\in R_{q_{L-1}}\}$ or simply $\{[m[X]]_{q_i}\}_{i\in L}$. It not only reduces the bit width of coefficients but also provides the possibility of parallel computation among different residue polynomials in $R_{q_i}$. All the computations on the polynomial level can be easily extended into the RNS-based residue polynomials.}

\begin{figure}
    \centering
    \includegraphics[width = 0.85\linewidth]{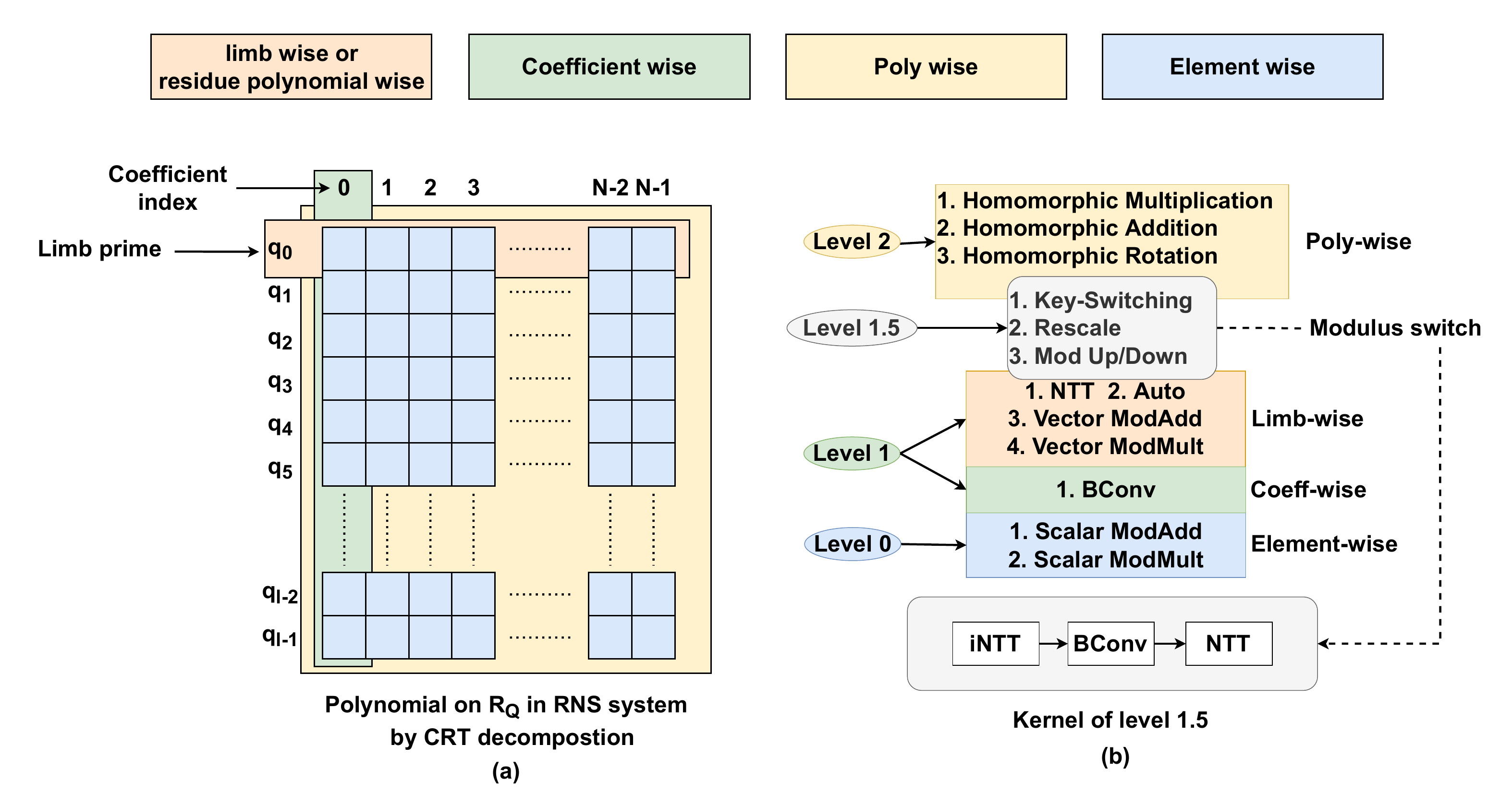}
    \vskip -2.5ex
    \caption{{\color{rebuttal}(a)The limb-wise and coefficient-wise data in a polynomial on $R_Q$ with RNS system. (b)The level of HE operations.}}
    \label{Level_op}
    
\end{figure}

{\color{rebuttal}The level of basic HE operations is shown in Figure \ref{Level_op}.b, which is categorized by the types of operands, which are represented as element-wise, residue-polynomial-wise, coefficient-wise, and polynomial-wise as shown in Figure \ref{Level_op}.a. 
For example, given two ciphertexts $\mathbf{ct_0}$$=(a_0\in R_Q,a_1\in R_Q)$ and $\mathbf{ct_1}$$=(b_0\in R_Q,b_1\in R_Q)$, the Homomorphic Addition (HADD) operates on their polynomials by computing $\mathbf{ct_{add}}$$=(a_0+b_0,a_1+b_1)$. When it is broken into residue-polynomial-wise vector Modular Addition (MADD), the two ciphertexts can be represented as $\mathbf{ct_0}$$=(\{[a_0]_{q_i}\}_{i\in L},\{[a_1]_{q_i}\}_{i\in L})$ and $\mathbf{ct_1}$$=(\{[b_0]_{q_i}\}_{i\in L},\{[b_1]_{q_i}\}_{i\in L})$. Then the HADD means performing $\mathbf{ct_{add}}$$=(\{[a_0]_{q_i}+[b_0]_{q_i}\}_{i\in L},\{[b_1]_{q_i}+[a_1]_{q_i}\}_{i\in L})$.
The computing kernel of level 1.5 and lower (level 1/0) is of importance in many architectures including EFFACT.}

\subsection{Number Theoretic Transformation (NTT)}
NTT is a variant of Discrete Fourier Transform (DFT), but it performs computation in a finite field. In CKKS schemes, NTT is used to speed up polynomial or residue polynomial multiplication. Performing polynomial multiplication in the out-of-NTT domain takes $O(N^2)$ asymptotic time, whereas the same operation only requires $O(NlogN)$ asymptotic time in the NTT domain, similar to the way FFT speeds up convolution. To eliminate the cost of padding to 2N-point NTT when performing N-point multiplication, \cite{nwc0, nwc1} proposes the Negative Wrapped Convolution (NWC) algorithm, which further improves the efficiency of NTT polynomial multiplication. {\color{rebuttal}The NWC-based NTT algorithm is defined as follows, let $A(X) = NWC\_NTT(a(X))$, we have:
\begin{equation}\label{eqn-2} 
A_j = \sum_{i=0}^{N-1}a_i\omega_{2N}^{(2i+1)j}\;\;\;\;mod\;\;Q
\end{equation}
where $A_j$ is the j-th coefficient of $A(X)\in R_Q$, $a_i$ is the i-th coefficient of $a(X)\in R_Q$, and $\omega_{2N}$ is the 2N-th root of prime $Q$. With the similar algorithmic optimization of FFT\cite{nwc0}, the equation can be broken down into a sequence of butterfly operations on $R_Q$.}

Due to the linearity and bit-reversal property of NTT operation\cite{nwc0, nwc1}, many arithmetic characteristics have been maintained between the NTT domain and out-of-NTT domain, we conclude these characteristics of NTT operation as follows: 
\begin{equation}\label{eqn-3} 
\begin{aligned}
NTT(a*b) &= NTT(a) \cdot NTT(b)\\
NTT(a+b) &= NTT(a)+NTT(b)\\
NTT(\sigma_s(a)) &= BR(\sigma^\prime_s(BR(NTT(a))))
\end{aligned}
\end{equation} 
where $a, b\in R_Q$, "$*$" is the convolution signal, "$\cdot$" means residue-polynomial-wise vector production, "+" means residue-polynomial-wise vector addition, "$\sigma_s()$" is the s-step automorphism operation in the out-of-NTT domain, "$\sigma^\prime_s()$" is the s-step automorphism operation in NTT domain, "BR" denotes the bit reversal operations.


\begin{figure*}[t]
      \centering
      \includegraphics[width=0.85\textwidth]{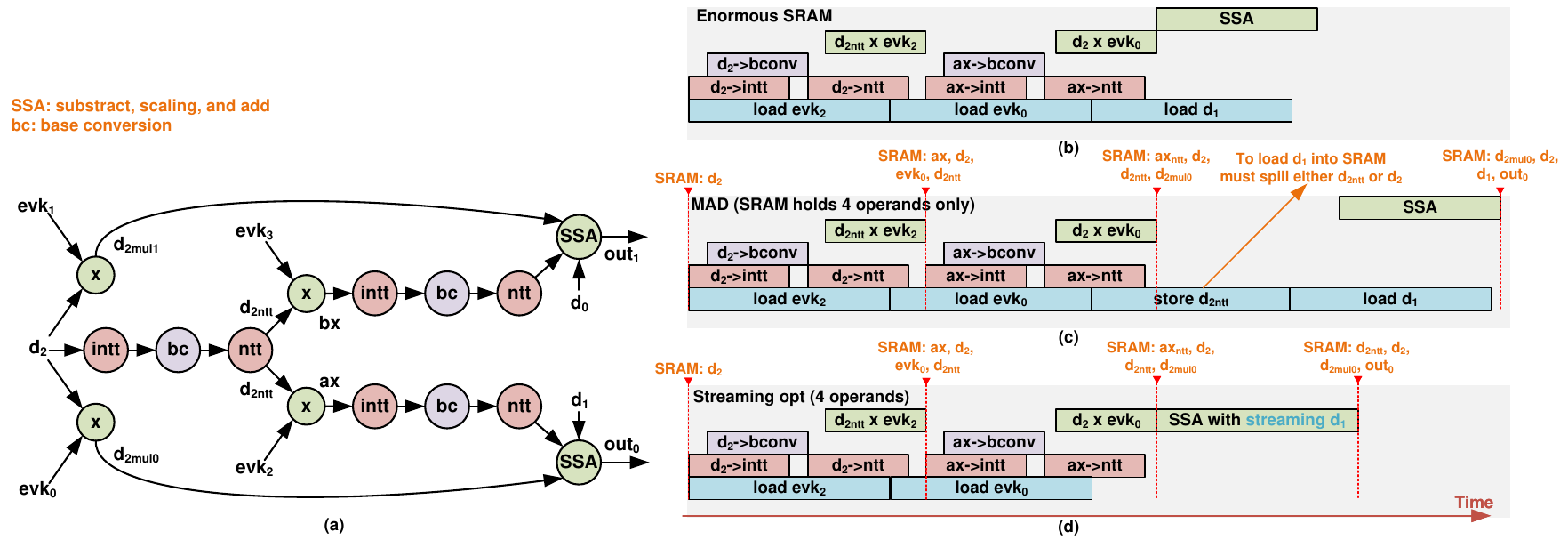}
       \vskip -2ex
    \caption{{\color{rebuttal}A toy example of performing key-switching in HMULT. $d_0$, $d_1$, and $d_2$ are the intermediate multiplication results noted in Section \ref{sec:bconv}, and the $evk_0$ and $evk_2$ are $evk_a$'s Q base and P base representations where the $evk_a$ is the component of evaluation key \textbf{evk}=$(evk_a,evk_b)$. We only show the timing diagram of the branch below the data flow graph (DFG). (a) Example key-switching DFG, (b) The timing graph of architectures with enormous SRAM and buffers that can hold all temporary operands, (c) The timing graph of MAD with limited SRAM and buffers that can only hold 4 operands, (d) The timing graph of MAD with our streaming optimization, in which we successfully reserve the $d_{2ntt}$ for the reuse in the branch above DFG, reducing extra spills. The latency of streaming optimized instruction is determined by the longest latency of the merged instructions. Here is the latency of loading $d_{1}$.}}
    \label{finegrain}
\end{figure*}
\subsection{Key-switching and Base Conversion (BConv)}\label{sec:bconv}
{\color{rebuttal}
Homomorphic Multiplication (HMULT) on two ciphertexts $\mathbf{ct_0}$$=(a_0\in R_Q,a_1\in R_Q)$ and $\mathbf{ct_1}$$=(b_0\in R_Q,b_1\in R_Q)$ requires both polynomial-wise multiplications and a key-switching process. It first computes the intermediate results $(d_0,d_1,d_2)$, where $d_0=a_0\cdot b_0$, $d_1=a_1\cdot b_0+a_0\cdot b_1$, and $d_2=a_1\cdot b_1$. The tuple $(d_0,d_1,d_2)$ is decryptable under the secret tuple $(1,s,s^2)$, where $s$ is the secret key.
The \textbf{evk}$=(evk_a,evk_b)$ is a ciphertext in the ring $R^{2}_{PQ}$ with a extended special prime $P$. The RNS can also be applied to the P base by $P=\prod_{i=0}^{k-1}p_i$. Such an \textbf{evk} is different from the input ciphertext's secret key, 
which is applied in key-switching to change the secret key back to $(1, s)$. The key-switching is the process by computing \textbf{ksw}=$P^{-1}(d_2\cdot$ \textbf{evk}), and the final HMULT result is $\mathbf{ct_{mult}}$=$(d_0,d_1)+$\textbf{ksw}.}

Base conversion is a major operation in the key-switching. It converts a set of residue polynomials from one modulus set to another set, therefore, it can change the intermediate results of HMULT to match the \textbf{evk}'s modulus set.
For a RNS representation $a_C$ on primes set $C = \{q_0, q_1, ..., q_{l-1}\}$, the fast base conversion from primes set $C$ to $B = \{p_0, p_1, ..., p_{k-1}\}$ is define as follows:
\begin{equation}\label{eqn-1} 
BConv_{C\rightarrow B}(a_C) = \{(\sum_{j=0}^{l-1}{(a_C[j]\cdot \hat{q_j}^{-1})_{q_j}\cdot \hat{q_j}})_{p_i}\}_{\quad 0\leq i<k}
\end{equation}
where $\hat{q_j} = \prod_{j^\prime \neq j}{q_{j^\prime}}$, $(\cdot)_{q_i}$ means reduce the number to $Z_{q_i}$. BConv operation is widely used in the CKKS scheme, almost as frequent as NTT/iNTT.

\subsection{Prior FHE accelerators}

In prior resource-unconstrained FHE accelerator designs, enormous on-chip SRAM and resource-hungry functional units, e.g., base conversion unit and NTT/iNTT unit, are used for tackling the memory-hungry and compute-hungry challenges in FHE computations. 
Despite their remarkable improvement, the use of hundreds of megabytes of SRAM and fully pipelined functional units makes the area and energy consumption high. Meanwhile, the off-chip memory bandwidth and functional units stay idle most of the time, e.g., less than 50\% HBM and $\sim$25\% base conversion utilization\cite{samardzic_craterlake_2022,sharp-kim}. Therefore, prior accelerators have a low efficiency with high cost due to the lack of memory and computing efficiency exploration.

{\color{revised}SHARP\cite{sharp-kim} provides an in-depth analysis of the impact of word length of slots on the security level and the data movement, thus, greatly enhancing area and power efficiency by exploring the optimal word length and a NoC-friendly hierarchical architecture. SHARP's insightful proposals can be extended to other accelerators such as ARK\cite{kim_ark_2022}, and are orthogonal to EFFACT as well.}
{\color{revised}MAD\cite{MAD} proposed a set of novel caching strategies to explore memory efficiency leading to enormous SRAM reduction. With the O($\alpha$) caching data path which only causes load/store of intermediate results at the modulus switching and uses computing resources' buffers to buffer intermediate results in other operations, DRAM transfers are reduced by 44\% compared to the naive bootstrapping baseline. However, these caching strategies are still within the category of custom manual adjustments, which only optimize several HE primitives and rely on the buffers on the computing resource side and large numbers of computing resources to hold and consume the intermediate results.
There is parallelism unexplored between the HE primitives when given limited SRAMs and buffers.
On the other hand, MAD requires large amounts of computing resources to immediately consume their data flow, or severe memory spills and data flow stalls arise. The computation inefficiency problem in the prior architecture is still unsolved.}
While SHARP and MAD have paved the way for efficient FHE accelerators, the incomplete exploration of memory efficiency and computing resource optimization makes them still fall short of achieving a highly efficient design with low cost.
\begin{figure}
    \centering
    \includegraphics[width = 0.75 \linewidth]{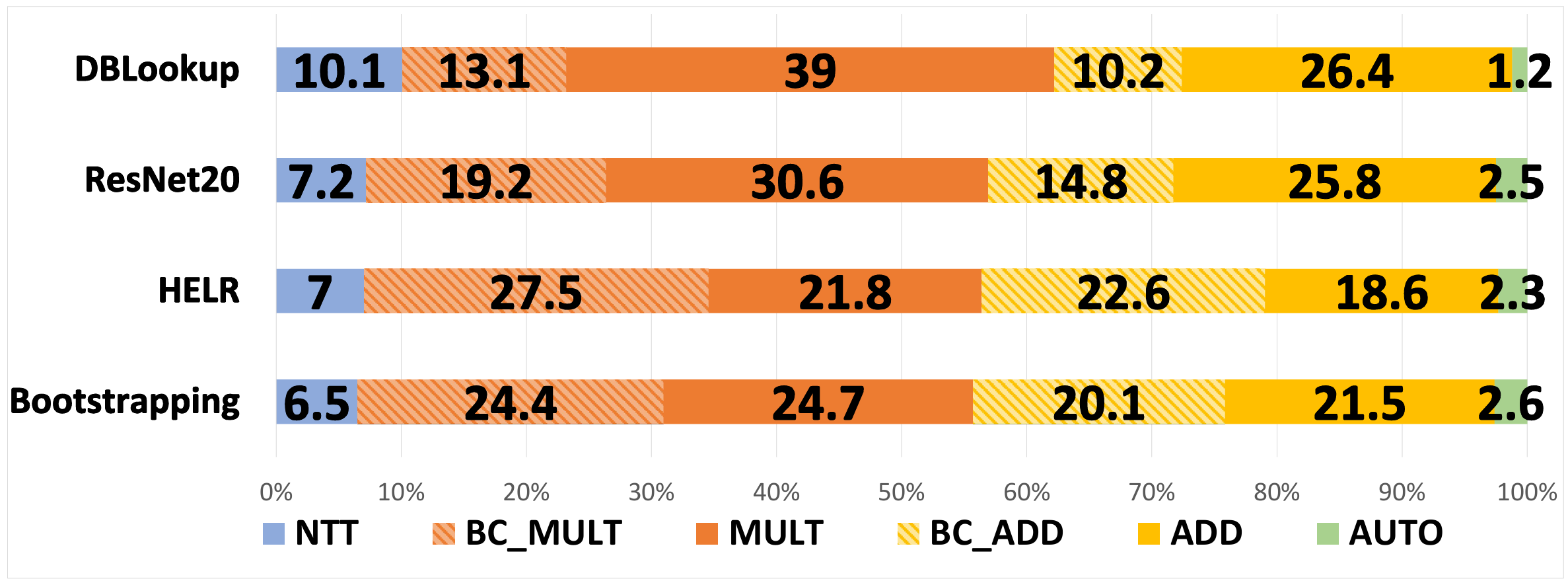}
     \vskip -2ex
    \caption{\color{rebuttal}Residue polynomial level instruction counts in DBLookup, ResNet20, HELR and Bootstrapping. BC\_MULT and BC\_ADD are MULT and ADD instructions used in BConv, while MULT and ADD represent the normal MULT and ADD except those in BConv.}
    \label{op_anal}
\end{figure}
\section{Opportunities and EFFACT proposals}\label{proposal}
{\color{revised}The first opportunity comes from the computing resource side. MAD\cite{MAD} keeps a similar computing resource distribution as resource unconstrained accelerators to get extreme performance of function units and smoothly conduct their manual data flow without memory spills or pipeline stalls. This causes their computing resources to occupy nearly 90\% area of MAD. 
MAD rarely explores circuit-level function unit reuse and computing resource optimization, which are essential for creating highly efficient and cost-effective designs.
Figure \ref{op_anal} shows the analysis of different instructions in IR format of HELR and fully-packed bootstrapping benchmark. 
The IR program is at the residue polynomial level, where NTT instructions only account for 7\% and 6.5\% of all instructions.
The majority of instructions are MULT and ADD operations, which totally account for 90.7\% and 90.9\% of all instructions, in which there are 52.7\% of MULT instructions and 51.6\% of ADD instructions are used for BConv. We call those MULT and ADD not for BConv the normal MULT and ADD.


{\color{revised} For the computations of a single ciphertext, we observe that among the normal ADD and MULT instructions (those not for BConv), nearly 77.6\% of them cannot be well hidden by the computing of $iNTT-BConv-NTT$ (dnum = 4, a practical parameter widely used in \cite{sharp-kim,kim_ark_2022,MAD}) in previous architectures\cite{kim_ark_2022,samardzic_craterlake_2022}. 
Such a situation arises since (1) 77.6\% of the normal ADD and MULT instructions, e.g., so-called MatMul1D, BlockMatMul1D in HElib\cite{halevi_design_nodate}, appear behind long $iNTT-BConv-NTT$ chains. Such a pattern also frequently shows up in bootstrapping's CtS/StC, ResNet's convolutions\cite{ResNet_20}, and HELR's gradient calculation\cite{han_logistic_2019}.
Due to the coefficient-wise aggregation nature of BConv, only the NTT at the end of the last chain or iNTT in the front chain (occupying only 0.88\%$\sim$1.75\% among all the instructions) can reveal parallelism with normal ADD and MULT instructions behind chains, (2) due to the inevitable data dependency caused by digit-decomposed key-switching algorithm\cite{cryptoeprint:2019/688} and hoisting rotation algorithm\cite{cryptoeprint:2020/1203} which need aggregation or automorphism in the last step, a majority of data from the NTT at the end of chains cannot be immediately used for subsequent normal multiplication or addition unless they undergo a series of aggregations, further blocking the parallelism. 
Greatly enhancing the units for normal MULT and ADD instructions may relieve this problem, however, it is out of EFFACT's scope.
Based on the above observation, we propose three optimizations as follows:}

\subsubsection{Removing BConv units} Since BConv MULT and ADD instructions cannot execute in parallel with most normal MULT and ADD instructions, we break BConv into a series of vector MULT and ADD instructions in the residue-polynomial-wise execution mode and use the normal MULT and ADD units to execute BConv, removing the specialized BConv units which occupy 66.7\% area of all computing units in CraterLake\cite{samardzic_craterlake_2022}. 
On the other hand, due to the residue-polynomial-wise execution mode, the computation inefficiency caused by ciphertext level change is also well solved.

\subsubsection{Reusing function units at circuit level} Since a large number of normal MULT and ADD instructions cannot run in parallel with NTT and iNTT, we devise a circuit-level NTT reuse scheme that leverages reconfigurable technology to intelligently reuse NTT units and modular multiplication units with no performance penalty. 
The butterfly units' modular multipliers and adders within the NTT units can be reused as MAC units to substitute consecutive normal MULT and ADD instructions, providing speed-up with low hardware overhead. 

\subsubsection{Fine-grained NTT units} Since nearly 17\% of the total instructions (see Figure \ref{op_anal}) cannot run in parallel with NTT/iNTT, we tend to adopt a fine-grained NTT pipeline similar to \cite{agrawal_fab_2022,yang_poseidon_nodate} instead of the fully-pipelined NTT. 
{\color{rebuttal}Unlike the fully-pipelined NTT unit in which each NTT pipeline stage possesses its own modular multiplier and adder, the fine-grained NTT unit tries to share the same modular multiplier and adder among all the NTT pipeline stages.}
Therefore, the fully-pipelined NTT\cite{four_step} runs at a better performance with a larger area than the fine-grained one. However, in resource-constrained scenarios, the fully-pipelined NTT design can easily disrupt the equilibrium between speed and hardware overhead. 
Using the proportion in Figure \ref{op_anal} and assuming other instructions can be perfectly executed in parallel with NTT and normal MULT and ADD without considering the DRAM fetches, the fully-pipelined NTT design can only achieve $\leq$2.7$\times$ performance enhancement compared to the fine-grained NTT design at the cost of $\geq$8$\times$ total computing resource consumption.
Therefore, implementing such fully-pipelined NTT units is not a better trade-off in the case of a highly efficient and cost-sensitive design.
}

{\color{revised}Another opportunity lies in the on-chip memory side. MAD uses a novel cache scheme to reduce the SRAM cost. However, it does not fully explore the memory efficiency due to its manual optimization and the need for computing resources' distributed buffers used for buffering the intermediate results. They keep prior designs' buffers. In their data flow, data are firstly fetched from DRAM to the SRAM and then flowed to the computing resources. The data will stay between computing resources and their buffers until the computation stage is finished. Such a data path works perfectly when the on-chip SRAM and buffers are large enough to hold all the operands and intermediate results for instructions like in BTS, ARK, and SHARP. However, as shown in Figure \ref{finegrain}(c), the reduced SRAM and buffer size causes extra spills and, thus, limit MAD's performance. 

To overcome the above challenge, we propose streaming data flow optimization as depicted in Figure \ref{finegrain}(d).
In our optimization, the function units can get their operands either from the SRAM or directly from the DRAM, which not only reduces the extra stores but also enhances EFFACT's DRAM bandwidth utilization.
Our compiler uses a static analysis pass to decide which instructions should get their operands from the DRAM.
}

\begin{figure}
    \centering
    \includegraphics[width = 1 \linewidth]{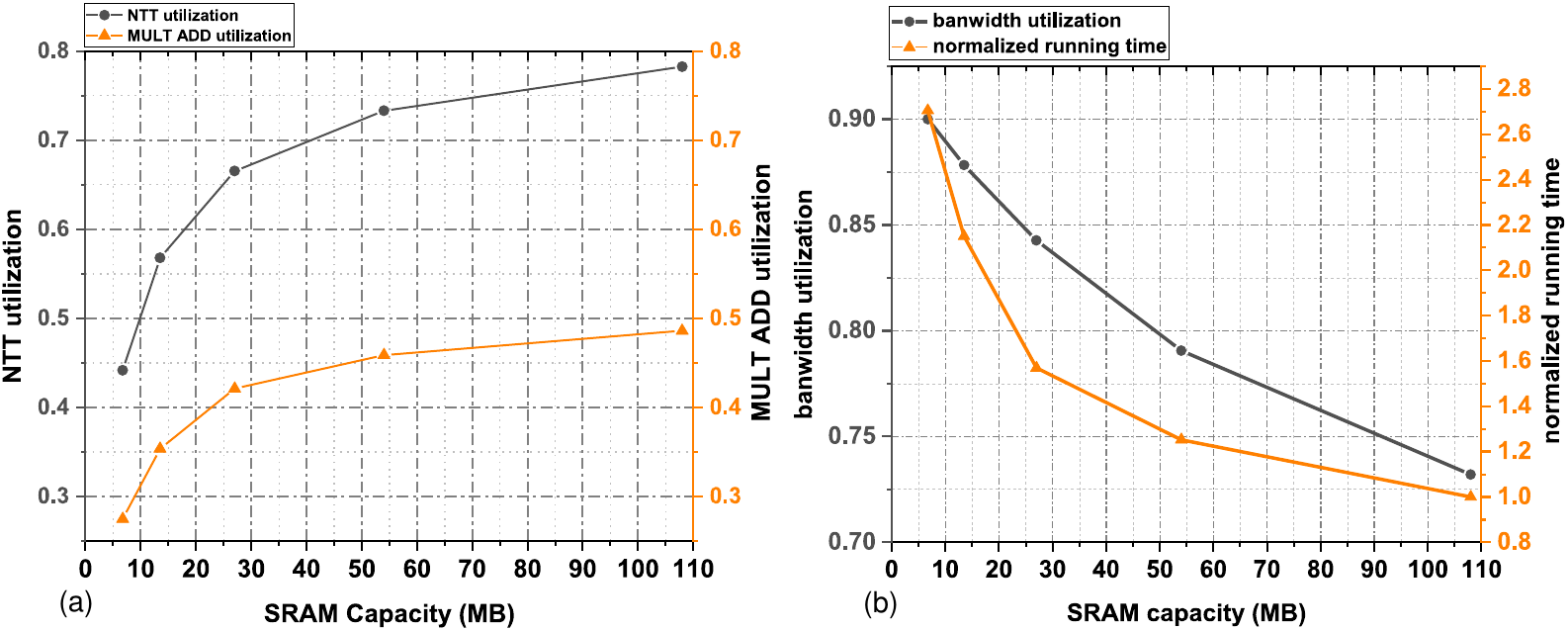}
     \vskip -2ex
    \caption{Impact of different SRAM sizes on utilization and total run time given certain computing resources, we do not show automorphism utilization since it is always low. (a) NTT, and MULT ADD units utilization with different on-chip memory, (b) DRAM bandwidth utilization and total running time with different on-chip memory.}
    \label{DRAM_SRAM}
\end{figure}

{\color{gg_r}Besides the above optimizations, EFFACT also provides an analysis of the interplay among SRAM capacity, bootstrapping running time, off-chip memory bandwidth utilization, and function unit utilization.}
The compiler of EFFACT automatically explores reuse opportunities and minimizes off-chip data movement at the program level using static scheduling and register allocation.  
Figure \ref{DRAM_SRAM} depicts the DRAM bandwidth, function unit utilization, and total runtime variation with different on-chip memory sizes in EFFACT, highlighting the performance and efficiency turning points at 27MB and 54MB. 
It is worth noting that the MULT and ADD units are almost saturated at $\leq$50\% because most of the normal MULT and ADD can only be executed serially with (i)NTT except for BConv operations, and in the EFFACT design, (i)NTT will also conduct some of the normal MULT and ADD.
We choose 27MB as a trade-off between performance, cost, and efficiency. We also use SimFHE \cite{MAD} to simulate DRAM transfers of MAD with O($\alpha$) caching strategy under identical architectural parameter setting, the result shows that the DRAM transfers of EFFACT are reduced by $\sim$40\% compared to MAD.

{\color{revised}To make EFFACT more practical and generic, we build a full-stack acceleration platform. Despite EFFACT's hardware design and optimizations, we also extract common residue polynomial level vector instructions from different FHE schemes to develop a generic ISA for EFFACT. At the same time, we also propose a compiler backend with code optimization for EFFACT that can be adapted to the recent compiler frontend\cite{HEAANMLIR}.}

\section{EFFACT Platform}

\subsection{ISA overview}

EFFACT looks into HE primitives and breaks those primitives at the level of residue polynomials. For generality, we analyze several FHE schemes' operations including CKKS, BGV, and BFV, and thus establish the vector ISA as shown in Table~\ref{table:isa}. For example, the MADD operation means adding one residue polynomial vector to another or a constant by a specific modulus. {\color{rebuttal}In the analyzed FHE schemes, operations of level 1 in Figure \ref{Level_op} fall into two categories. One is the residue-polynomial-wise operation in (i)NTT/automorphism or between two polynomials (feeding one polynomial's residue polynomial to the other polynomial's residue polynomial). The other is the coefficient-wise operation within one polynomial. Our ISA can fully support both categories. Our residue-polynomial-wise ISA naturally supports the residue-polynomial-wise operations between two polynomials. Since the coefficient-wise operations' behavior is the same for all the slots in the same residue polynomial, the coefficient-wise operations within one polynomial can be conducted by keeping a residue polynomial of the given polynomial in the memory and then feeding the next residue polynomial to it. In this way, the coefficient-wise operation is changed into the residue polynomial vector operation.} In addition, we keep some of the int64 operations (scalar subset) to perform control flow such as loops.

\begin{table}[h!]

\caption{EFFACT ISA}
 \vskip -2ex
\centering
\resizebox{\linewidth}{!}{
\normalsize
\begin{tabular}{c|c}
\hline
\textbf{Instruction} & \textbf{Description}\\
\hline
\hline
MMUL dest src0, src1(imm), modulus 
    & perform modular multiplication on residues \\
\hline
MMAD dest src0, src1(imm), modulus
    & perform modular addition on residues \\
 \hline
(i)NTT dest src0, modulus
    &  perform (i)NTT on a residues\\
\hline  
AUTO dest src0, imm, modulus
    & perform automorphism on a residues \\
 \hline
LoadRes dest, srcaddr
    & load a residue from main memory \\
 \hline
StoreRes destaddr, src0 
    & store a residue into main memory \\
 \hline 
 VecCopy dest, src0 
    & move residue among on-chip SRAM \\
 \hline 
 Scalar subset 
    & support loop, branch, and address calculation \\
 \hline
\end{tabular}
}
\label{table:isa}
 \vskip -2ex
\end{table}
\subsection{Compiler Design} \label{sec:compiler}
Our compiler optimizes the extended LLVM Intermediate Representation (IR) file\cite{LLVM} and generates the executable machine program while also partially supporting control flow features such as loops and branches. 

\subsubsection{Code Optimizations}
Our compiler begins by parsing the IR file and the hardware description, lowering the IR instructions to EFFACT's ISA while maintaining the Static Single Assignment (SSA) form. Then it performs copy propagation and constant propagation to eliminate redundant vector copies across different on-chip SRAMs and reduce constant calculation during execution. Our compiler also employs partial redundancy elimination (PRE) using the algorithm described in  \cite{kennedy_partial_1999,knoop_lazy_1992,briggs1994effective} to eliminate the code redundancy. 
Additionally, our compiler performs computation merge as mentioned in Section \ref{sec:arch} through a peephole optimization pass.
Code optimization is crucial as the automatic IR translator introduces some redundant code. In fully-packed bootstrapping, our code optimizer eliminates 12.9\% of instructions, a task that was previously done manually in prior works\cite{kim_ark_2022,MAD,sharp-kim}. Given that previous designs relied on manual optimizations, we have excluded code optimization from our evaluation.



\subsubsection{Static Scheduling \& Memory Allocation}
An alias analysis \cite{Andersen2005ProgramAA} is first performed before scheduling instructions to chain the load/store which may point to the same address in the correct order. Then we schedule the SSA-formed instructions globally as described in \cite{porpodas_caesar_2013, abraham_efficient_2000, sweany_dominator-path_1992} to get the optimal parallelism. 
One big challenge in the compiler is how to manage the on-chip SRAM and the HBM stack to minimize spilling and reduce the load/store when allocating SRAM. Since our vector operation is at the residue polynomial level, we can split the on-chip SRAM into several parts which are the size of one or two residue polynomials, and view each part as a register. Thus, the linear register allocation algorithm \cite{wimmer_linear_2010,poletto_linear_1999,traub_quality_1998} can be adopted to allocate on-chip SRAM and manage the HBM. 

\subsubsection{Instruction Merging for Streaming Memory Access}
The compiler identifies load operations with a single consumer and merges them as a new streaming operation. The original load operations will not be considered in the succeeding memory allocation phase. The same technique also works in store operations and 
between different function units. 

\subsubsection{Machine Code Generation}
Ultimately, the binary program is obtained by translating the optimized IR into EFFACT's machine code.

\subsection{Streaming Memory Access}\label{sec:stream}

In EFFACT architecture, there is a stream memory controller to manage memory accesses. 
Streaming is a pattern that involves the continuous flow of data through computing units and DRAM.
EFFACT employs partial streaming memory access \cite{somogyi2006spatial,nowatzki2017stream}, in which part of data can flow directly from DRAM to function units upon arrival instead of waiting for the SRAM chunk to be filled. 
It aims to maximize computing resource utilization, reduce memory latencies, and more essentially, relieve the SRAM pressure. 
For stream implementation, a separate FIFO address space is added to efficiently collect data. The memory controller handles concurrent accesses from HBM to the on-chip SRAM or among on-chip SRAM and the aforementioned FIFO space. High parallelism with low overhead can be achieved by fully utilizing the associated resources through arbitration of memory accesses. 

As mentioned in Section \ref{sec:compiler}, load operations or normal functional operations with only one consumer will be processed with instruction merging and are allowed to use streaming memory access. 
The merged instruction will first issue the first operation it merges.
Through streaming, vector data that is returned by the first operation can be directly read into the aforementioned FIFO space one by one and subsequently dispatched straight to the second operation's FU or the DRAM. If there are no other data dependencies, the second operation will execute upon arrival of the data without needing to wait for the preceding operation to complete. The data is processed on arrival, eliminating the requirement of waiting for the on-chip SRAM to fill up, and enabling nearly real-time processing of data, as shown in Figure \ref{finegrain}(d).  
Otherwise, in the case where data will be used by multiple consumers, it is filled into the on-chip SRAM for reuse.


\subsection{EFFACT Architecture} \label{sec:arch}
\begin{figure}
    \centering
    \includegraphics[width = 0.75 \linewidth]{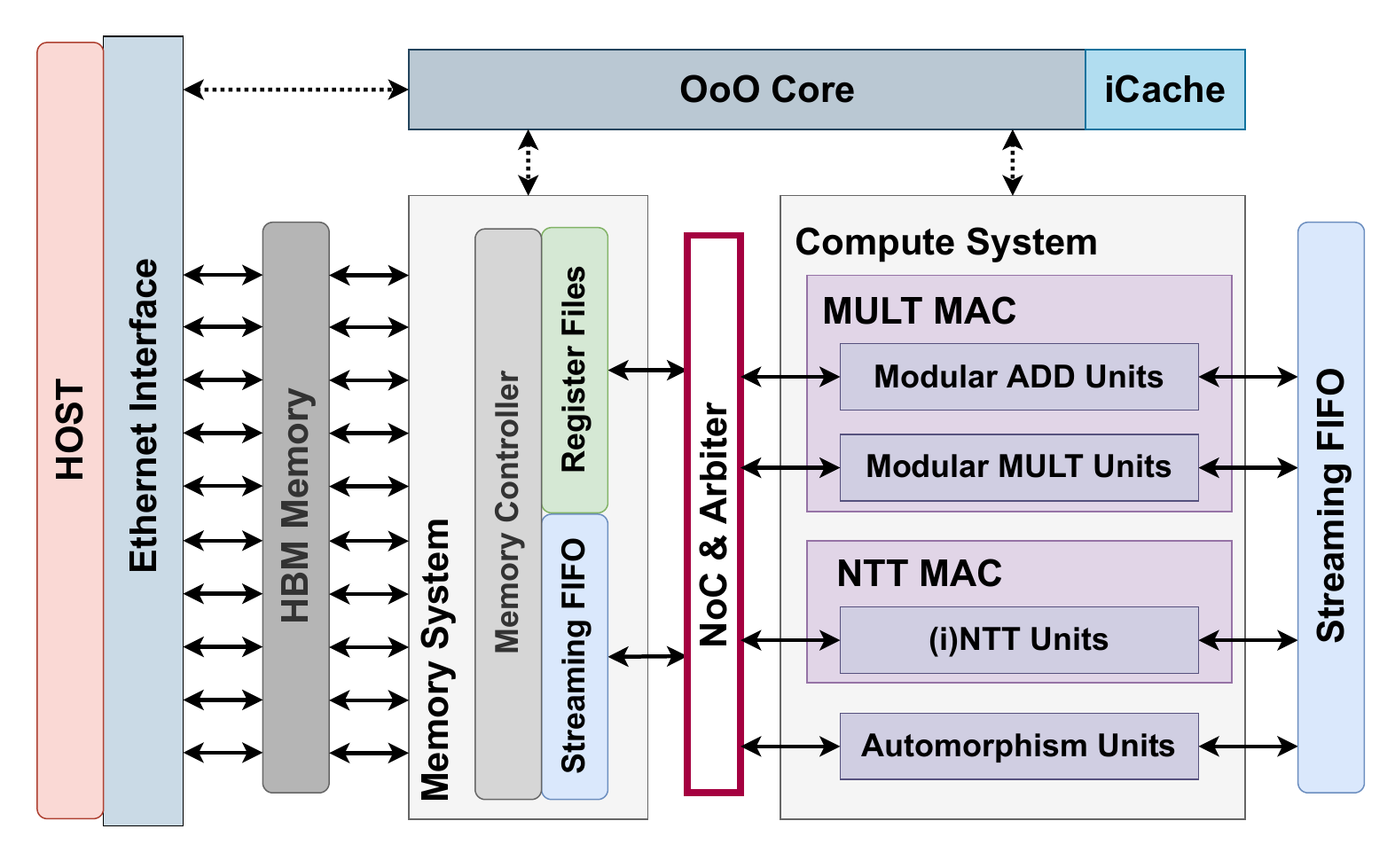}
     \vskip -2ex
    \caption{EFFACT overall hardware architecture.}
    \label{fig:arch}
\end{figure}
EFFACT architecture consists of an Out-of-Order (OoO) control core, four functional units (ModAdd, ModMult, NTT, and Auto), FIFOs, register files (RFs, implemented using SRAM), and the interface with off-chip HBM. This section will introduce how these components work in our architecture in Figure \ref{fig:arch}. 

\subsubsection{OoO core}
Both the SRAM and streaming FIFO will request data from DRAM. To fully utilize the DRAM bandwidth, we allow the streaming FIFO and SRAM to compete for DRAM transferring. Since the instruction merge may chain load/store with the fine-grained NTT which is slower than DRAM transferring, tying DRAM to NTT will lead to underutilization of DRAM bandwidth. Therefore, competing for DRAM transferring becomes necessary. 
To manage the competition of DRAM requests, we introduce an OoO controller that issues instructions when function units are temporarily available and controls in-flight instructions.
We implement an OoO core using the scoreboard to track the dependencies and availability of instructions.
Since memory ordering has been enforced by the compiler, the results of instructions are ensured to be written back to the register file or the memory in the correct order.

\subsubsection{Memory and Interfaces}
The on-chip SRAM is connected to a multi-channel off-chip HBM for staging data in order to enable increased data reuse and reduced access overhead, as mentioned in Section \ref{sec:stream}. Also, the partial streaming memory access scheme allows direct communication between the function units and HBM to save the step of filling the on-chip SRAM. A memory controller and an arbiter are responsible for managing the data access between function units and the SRAM via banks. An HBM controller manages the load/store operations of the HBM.

Since the SRAM and computing resources are largely decreased thanks to the proposed streaming and hardware optimizations, it alleviates the stringent requirement for large on-chip SRAM bandwidth. Therefore, we are able to reduce the on-chip bandwidth to further reduce the area and energy consumption.
Unlike MAD, ARK, and CraterLake, which require at least 90TB/s on-chip SRAM bandwidth, the streaming access and smart function unit reuse scheme relieves the pressure on the SRAM ports and thus, EFFACT only requires about 30TB/s on-chip SRAM bandwidth, giving a roughly threefold improvement.


\subsubsection{Reconfigurable NTT units} 
\begin{figure}
    \centering
    \includegraphics[width = 0.8\linewidth]{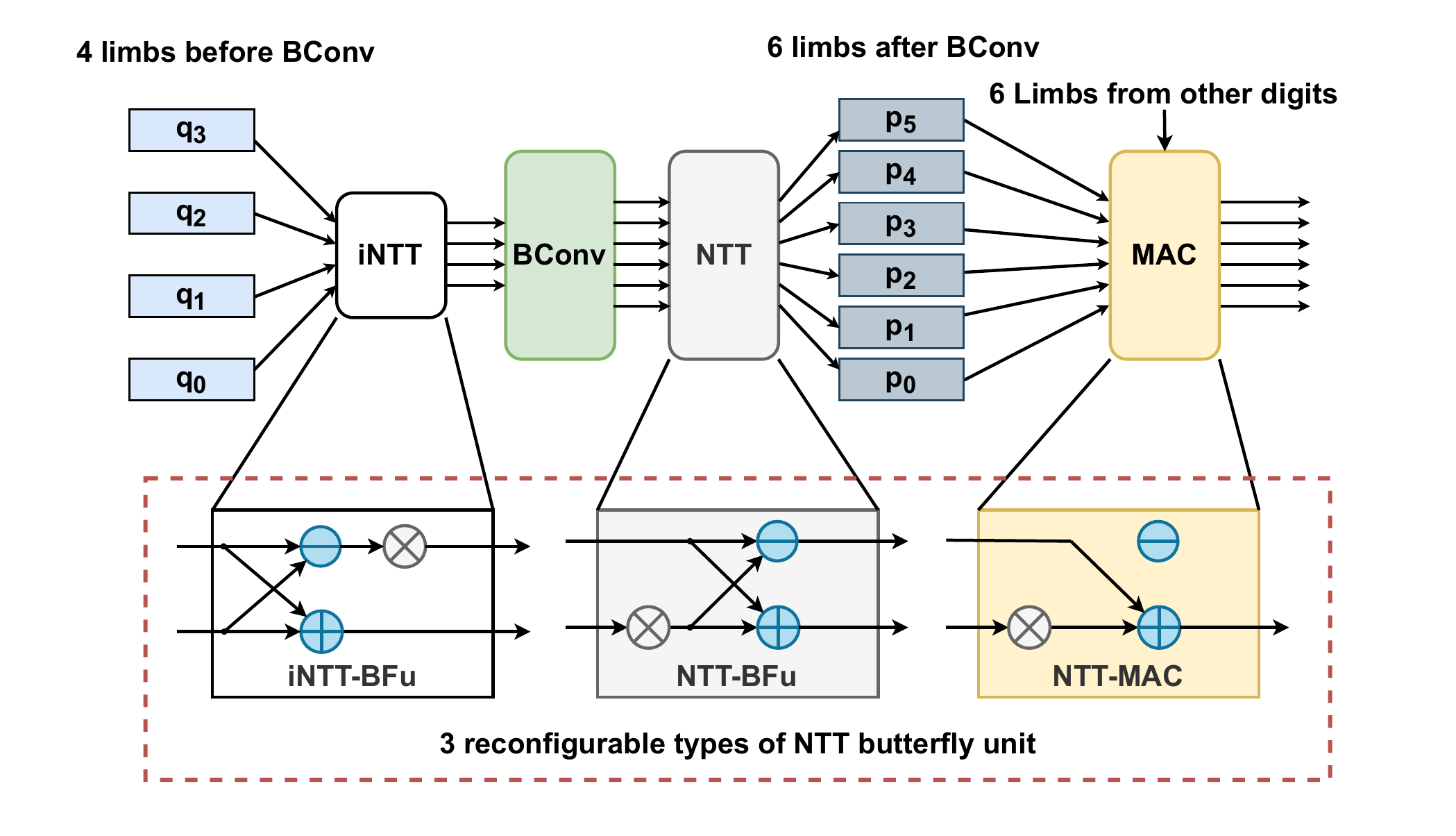}
     \vskip -2ex
    \caption{NTT reuse scheme to accelerate MAC operation.}
    \label{BConv_impl}
\end{figure}

{\color{rebuttal}The baseline fine-grained NTT architecture of EFFACT is similar to \cite{agrawal_fab_2022}. 
However, EFFACT makes several improvements to better adapt to a vector accelerator. 
First, EFFACT employs the CG-NTT\cite{CGNTT} algorithm to implement the fine-grained NTT units due to their vector-friendly characteristics instead of the Cooley-Tukey in \cite{agrawal_fab_2022}. 
Second, we diminish the bit-reversal operation of each coefficient vector. We perform the bit-reversal operation on twiddle factors rather than the N coefficients since the bit-reversal operation of twiddle factors (TFs) can be done with much less cost during TFs seed pre-computation and TF on-the-fly generation. 
Moreover, we notice that some modular multiplications and additions cannot run in parallel with the NTT and the NTT naturally possesses a mult-accumulate (MAC) data path in its circuit implementation as shown in the middle of Figure \ref{BConv_impl}. Therefore, we propose to reuse the NTT as a MAC unit while \cite{agrawal_fab_2022} cannot.
}

{\color{rebuttal} Figure \ref{BConv_impl} shows a sequence of $iNTT-BConv-NTT$ followed by normal MULT and ADD, which shows up frequently in operations such as matrix multiplications, convolutions, and hoisting rotations. Based on the observations proposed in Section \ref{proposal}, EFFACT reuses NTT unit under following execution modes. 
First, the butterfly units at the bottom of Figure \ref{BConv_impl} can be reused as both NTT-butterfly units and iNTT-butterfly units (the middle and left) just like ARK\cite{kim_ark_2022} by adding several multiplexers to change the position of the multiplier in the data path. Second, EFFACT also supports configuring the NTT data path as the MAC units to accelerate those MAC operations that can not run in parallel with NTT. It is implemented by simply masking the subtraction out of the data path and picking the MAC result out of the original NTT pipeline. Through this reconfigurable technology, EFFACT gets performance enhancement in MAC operations without much hardware overhead.}

\subsubsection{Auto units}
Automorphism is a sub-operation of HROT, where the automorphism with $s$-step is defined as $\sigma_s(\cdot)$. It can be regarded as a straightforward process of address mapping and sign transformation. In address mapping, the new index of position $i$ for rotation step $s$ is calculated as follows:
\begin{equation}\label{eqa}
\begin{aligned}
index_{new}(i, \;s) = i*5^s\;mod\;N
\end{aligned}
\end{equation}
Inspired by ARK\cite{kim_ark_2022}, we discovered that data in each row remains in the same row after the automorphism operation. 
This allows us to decompose the $2^{16}$-element permutation into $N_L$-element permutations inside each row and design the auto-mapping units with sign transformation, where $N_L$ is the number of lanes in EFFACT. 
Such units have also been used in ARK\cite{kim_ark_2022} to vectorize automorphism. 

\begin{figure}
    \centering
    \includegraphics[width = 0.8 \linewidth]{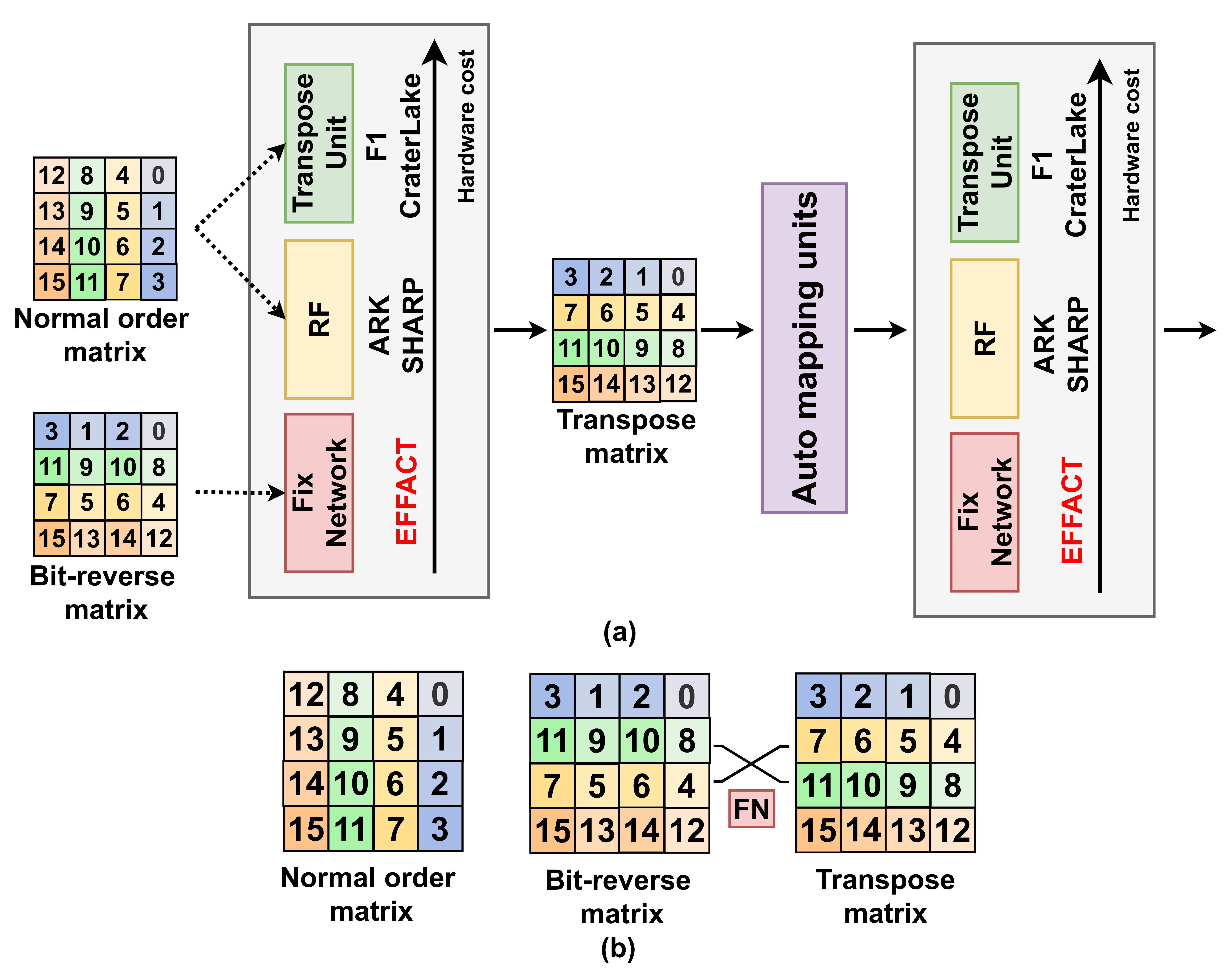}
     \vskip -2ex
    \caption{(a) Hardware structure of different automorphism units and hardware cost comparison, (b) An example of how bit-reverse character works on the 16-point matrix transposition.}
    \label{fig:autounit}
    \vskip -2ex
\end{figure}

{\color{revised} However, for the transpose before and after the auto-mapping units in Figure \ref{fig:autounit}(a), SHARP\cite{sharp-kim} and ARK\cite{kim_ark_2022} transpose the input matrix by accessing different rows of their register files simultaneously, which requires their RFs heavily banked. 
This is not suitable for a vectorized accelerator like EFFACT which can only access SRAM row by row.
On the other side, the transpose units proposed by CraterLake\cite{samardzic_craterlake_2022} demand excessive global connections, causing a large area.
Therefore, EFFACT proposes a modified algorithm replacing the expensive matrix transpose operation with a fixed network (FN). 
We notice when the coefficients in the NTT domain follow a bit-reversal order, each row of the coefficients matrix shares the same transform pattern to turn to the transposed matrix as shown in Figure \ref{fig:autounit}(b), while the transformation in the column can be conducted by changing the SRAM fetching order.}
This implies that we can simply leverage a fixed network to implement the comprehensive transposed operation. 


\subsubsection{Merge computation into BConv}\label{sec:Merge_compute}
{\color{rebuttal} Prior works\cite{samardzic_f1_2021, kim_ark_2022} uses the Montgomery modular multiplier due to the relatively lower hardware overhead. However, it will bring extra Montgomery representation transformation penalties when dealing with the modulus switching operations. Meanwhile, the post-processing of iNTT will also bring one extra modular multiplication with a constant. To eliminate both issues, EFFACT analyze the widely-used $iNTT-BConv-NTT$ computation flow and propose to merge these extra computations into the BConv.}

{\color{rebuttal} 
In classic Montgomery algorithm \cite{Montgomery1985ModularMW}, input data will be converted into their single-Montgomery (SM) representations like $X\rightarrow XR\; mod\;Q$. The Montgomery modular multiplication is computed on SM representations of input data by $MontMult(XR,\;YR,\;Q)=XYR\;mod\;Q$, which maintains the same representation between the input and output data. 
The data in EFFACT will be maintained in their SM representations through the whole process when no modulus transformation happens. 
However, operations like rescale and key-switching require data in the non-Montgomery (NM) format to perform modulus transformation, causing extra Montgomery representation transformation penalties. 
To reduce the extra Montgomery transformations, we propose a double-Montgomery (DM) representation of constant numbers defined as $X\rightarrow XR^{2}\; mod\;Q$. Then the DM representation can help to merge the Montgomery transformation into BConv as follows. 

Initially, the $(\hat{q_j}^{-1})_{q_j}$ in equation \ref{eqn-1} is kept in the NM representation. The input data $a_C[j]$ in equation \ref{eqn-1} naturally stays at the SM representation. Therefore, the Montgomery multiplication result of $a_C[j]$ and $(\hat{q_j}^{-1})_{q_j}$ will be in the NM representation. 
To this end, we pre-compute the constant $(\hat{q_j})_{p_i}$ in equation \ref{eqn-1} in the DM representation. The Montgomery multiplication results with $(\hat{q_j})_{p_i}$ in the DM representation will directly change the NM-represented intermediate result from the last multiplication back into the SM representation. 
With this optimization, we eliminate the Montgomery representation transformation in the modulus transformation.}

On the other hand, a final constant multiplication with $\frac{1}{N}$ is needed when finishing iNTT butterfly operations. 
In the $iNTT-BConv-NTT$ computation flow, each iNTT operation is followed by a BConv operation, giving the opportunity to merge the $\frac{1}{N}$ constant multiplication into the first constant multiplication of BConv operation, i.e. rewrite the constant $(\hat{q_j}^{-1})_{q_j}$ in equation \ref{eqn-1} as $(\hat{q_j}^{-1}*\frac{1}{N})_{q_j}$ which can be pre-computed.

In total, the computation merge that eliminates both the post-processing of iNTT and the Montgomery representation transformation penalties can be solved by redefining the right side of BConv in equation \ref{eqn-1} as:

\begin{equation}\label{eqf}
\begin{aligned}
    \{(\sum_{j=0}^{l-1}{(a_C[j]^{SM}\cdot(\hat{q_j}^{-1}*\frac{1}{N})^{NM})_{q_j})\cdot \hat{q_j}^{DM}})_{p_i}\}_{0\leq i<k}
\end{aligned}
\end{equation}


\section{Experimental Methodology}

We evaluate a complete EFFACT system: the compiler is built using C++ with LLVM-like IR and passes, and the microarchitecture is fully implemented in RTL and synthesized by using LVT TSMC 28nm technology node by Synopsis Design Compiler with a commercial SRAM IP\cite{28SRAM} licensed by TSMC. Moreover, we further synthesized our RTL design to the Xilinx VCU128 evaluation board using Vivado 2021.1. 

\begin{figure}
    \centering
    \includegraphics[width = 0.4 \linewidth]{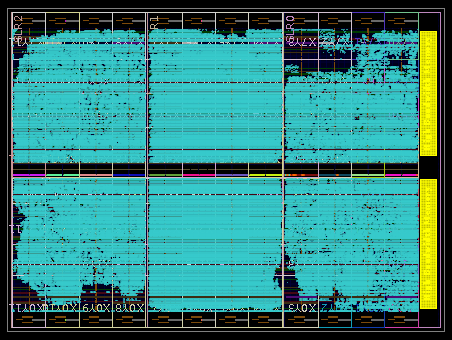}
     \vskip -2ex
    \caption{FPGA layout of EFFACT with 64 lanes.}
    \label{fig:layout}
    \vskip -2ex
\end{figure}


\subsection{Benchmarks and Parameters}
\noindent\textbf{Fully-packed bootstrapping:}
Bootstrapping is the pivotal operation of the FHE schemes, we compare the bootstrapping performance of EFFACT with the SOTA
GPU\cite{jung_over_nodate}, FPGA\cite{agrawal_fab_2022,yang_poseidon_nodate}, and ASIC\cite{samardzic_craterlake_2022,kim_ark_2022,kim_bts_2022} implementation. Our fully-packed bootstrapping algorithm consumes $L_{boot} = 15$ level, which includes 4 levels for the CtS procedure, 3 levels for the StC procedure, and 8 levels for the EvalMod procedure.

\noindent\textbf{Logistic regression:}
Logistic regression is a general model in machine learning which used for binary classification. Previous work HELR \cite{han_logistic_2019} proposed a highly efficient logistic regression algorithm based on the CKKS scheme, we regard this work as the baseline, and target at the training phase. The experimental results show that after 30 iterations of training in EFFACT, the accuracy of the inference phase can reach 96.67\%.

\noindent\textbf{ResNet-20:}
ResNet-20\cite{ResNet_20} implemented the ResNet-20 DNN model based on the CKKS scheme, which consists of numbers of bootstrapping and homomorphic convolutions, we evaluate the performance of the inference on a single encrypted image as in the original implementation.

\noindent\textbf{DB-lookup based on BGV:}
As mentioned earlier, EFFACT is a universal acceleration platform for BGV, BFV, and CKKS. We also evaluate the performance of the DB Lookup application proposed in HElib \cite{halevi_design_nodate} based on the BGV scheme, and we take F1's implementation as our baseline.

In order to show it more intuitively, we list the parameters used in the fully-packed and 256-slot bootstrapping in Table \ref{boots_param}. HELR starts computations at level 23 and performs 256-slot bootstrapping per two iterations, 
The parameters of other benchmarks are the same as their original implementations. 

\begin{table}[h!]
    \centering
    \caption{Bootstrapping parameters}
    \vskip -2.5ex
    \scalebox{0.75}{
    \begin{tabular}{|c|c|c|c|c|c|c|c|c|}
        \hline
        $\#Slots$ & \textbf{$N$}& \textbf{$L$}& \textbf{$L_{boot}$}& \textbf{$L_{CtS}$} & \textbf{$L_{EvalMod}$} & \textbf{$L_{StC}$} & \textbf{$log(q)$} & \textbf{$dnum$} \\
        \hline
        $2^{15}$ & $2^{16}$ & 24 & 15 & 4 & 8 & 3 & 54 & 4 \\
        \hline
        $2^8$ & $2^{16}$ & 24 & 13 & 3 & 8 & 2 & 54 & 4 \\
        \hline
    \end{tabular}}
    
    \label{boots_param}
\end{table}

\subsection{Compiler}
EFFACT's compiler passes are LLVM-style and are in line with the EFFACT ISA. We implemented passes for code optimization, memory ordering, and static scheduling. We also implemented code generation for the EFFACT microarchitecture. The compiler is run on a workstation with 4 Intel Xeon 3.40 GHz CPUs. 


\subsection{System-Level Integration \& Experimental Platform}
EFFACT is implemented completely in RTL, including an OoO core and four function units, together with a User Datagram Protocol (UDP) unit. We provide an ASIC version featured with 1.2-TB/s HBM bandwidth and 1024 lanes which is synthesized without the UDP unit and HBM controller. The power and area of HBM are estimated using \cite{HBM}. We also implemented the RTL on the Xilinx VCU128 evaluation board as EFFACT's experimental platform.

{\color{gg} In the FPGA experimental platform, the evaluation board is connected to the PC through a 1000-Mbit Ethernet interface. We developed a C++ source code based on the socket library to communicate with the FPGA using the UDP. Data for each application begins off-chip and is loaded from FPGA HBM.
On the FPGA side, there is a corresponding module that unpacks the UDP-formatted data and deposits it sequentially into the HBM, and the results of the FHE program are also packed up and transmitted using the UDP. Furthermore, the functionality of EFFACT has been thoroughly verified by comparing it with Lattigo\cite{mouchet_lattigo_2020}.}


Figure \ref{fig:layout} demonstrates the layout of our design. Worth mentioning that the system in FPGA runs only at 12.5 MHz with 64 lanes, although we successfully synthesized it at 300 MHz with 256 lanes on Vivado. The major bottleneck lies in the routing congestion in which the congestion level reaches 7. The HBM bandwidth is also lowered through an asynchronous FIFO to ensure that we can correctly scale the performance of the 12.5 MHz with 64 lanes version to our target 300 MHz with 256 lanes FPGA-EFFACT and 1024 lanes ASIC-EFFACT.

\section{Evaluation}


  

\subsection{Area and Power Analysis}
\vskip -3ex
\begin{table}[h!]
  \centering
  \caption{ASIC-EFFACT breakdown}
   \vskip -2ex
  \small
  \scalebox{0.8}{
  \begin{tabular}{|c|c|c|}
    \hline
    \textbf{Components} & \textbf{Area($mm^2$)} & \textbf{Power(W)} \\
    \hline
    NTTU & 37.13 & 21.16  \\
    \hline
    MADDU & 3.59 & 3.51  \\
    \hline
    MMULU & 18.21 & 10.12 \\
    \hline
    AUTOU & 4.65 & 4.88 \\
    \hline
    SRAM & 81.50 &43.14\\
    \hline
    HBM & 29.60 &31.80\\
    \hline
     Others & 37.20 & 21.13 \\
     \hline
  \end{tabular}}
  
  \label{ASIC_break}
\end{table}
Table \ref{ASIC_break} shows the area and power breakdown of ASIC-EFFACT.
We adopt a fine-grained NTT unit in contrast to other ASIC designs and there are no additional transpose units and twisting units in the data path, which leads to only 11$\%$ area of the NTTU in ARK. Unlike other designs that use more than 200-MB SRAM, ASIC-EFFACT only requires 27 MB, dropping the SRAM area to 6$\%$ compared to ARK. In total, the SRAM occupies 38.46$\%$ area and 31.79$\%$ power with 30$\%$ and 29.22$\%$ going to FUs.

\vskip -2ex
\begin{table}[h!]
  \centering
  \caption{ASIC Resource Comparison}
   \vskip -2ex
  \small
  \scalebox{0.8}{
  \begin{tabular}{|c|c|c|c|c|}
    \hline
    \textbf{} & \textbf{Tech} & \textbf{Freq($GHz$)} & \textbf{Area($mm^2$)} & \textbf{Power($W$)}\\
    \hline
    F1 & 14/12 nm & 1-2 & 151.4 & 180.4 \\
    \hline
    BTS & 7 nm & 0.3-1.2 & 373.6 & 133.8 \\
    \hline
    CraterLake & 14/12 nm & 1-2 & 472.3 & 320.0 \\
    \hline
    
    ARK & 7 nm & 1 & 418.3 & 281.3 \\
    \hline
    CL+MAD-32 & 14/12 nm & 1 & 333.9 & 213.4 \\
    \hline
     \textbf{ASIC-EFFACT} & 28 nm & 0.5 & 211.9 &  135.7 \\
     \hline
  \end{tabular}}
  
  \label{ASIC_resourse}
\end{table}

Table \ref{ASIC_resourse} shows the resource comparison with recent ASIC designs. By technology scaling\cite{16sram,7sram,14global} (HBM keeps unchanged when scaling), ASIC-EFFACT only requires 0.783$\times$, 0.153$\times$, 0.257$\times$, 0.137$\times$, and 0.414$\times$ area consumption compared to F1\cite{samardzic_f1_2021}, BTS\cite{kim_bts_2022}, CraterLake\cite{samardzic_craterlake_2022}, ARK\cite{kim_ark_2022}, and CL+MAD-32\cite{MAD}. Meanwhile, ASIC-EFFACT nearly achieves the lowest absolute power among these designs. The main area and power drop come from the significant reduction in SRAM capacity (nearly identical to MAD, 2$\times$ compared to F1, more than 8$\times$ in others), no computing resources' buffers (some necessary pipeline registers in which one register is only lanes$\times$1 large) and the smart function unit reuse scheme. In the next subsection, we will demonstrate that ASIC-EFFACT reaches a higher performance density\cite{scale-out-processor} and power efficiency than prior ASIC designs even though we reduce the function units and SRAM capacity.

\vskip -2ex
\begin{table}[h!]
  \centering
  \caption{FPGA Resource Comparison}
   \vskip -2ex
  \small
\scalebox{0.72}{
  \begin{tabular}{|c|c|c|c|c|c|c|}
    \hline
    \textbf{Work} & \textbf{Platform} & \textbf{LUT} & \textbf{FF} & \textbf{BRAM} & \textbf{URAM} & \textbf{DSP}\\
    \hline
    FAB & Xilinx U280 & 899K & 2073K & 3840 & 960 & 5120 \\
    \hline
    Posidon & Xilinx U280 & 728K & 915K & 2048 & - & 8640 \\
    \hline
    \textbf{FPGA-EFFACT} & Xilinx VCU128 & 1246K & 2096K &  1343&  864&8212 \\
    \hline
  \end{tabular}
  }
  
  \label{FPGA_resourse}
\end{table}
  

Table \ref{FPGA_resourse} lists the resource utilization of our FPGA-EFFACT. Although FPGA-EFFACT only requires 7.6-MB SRAM, the BRAM and URAM utilization reaches more than 50$\%$ in the VCU128. This is because the BRAM and URAM array in the FPGA have a depth of 1024 and 4096, in which our residue polynomial mapping only uses 256 rows leading to more than $75\%$ rows unused.
{\color{yi}Meanwhile, to route FPGA-EFFACT, we use the routability strategy of Vivado which increases our LUT usage from $\sim$900K by default to 1246K.}

\subsection{Performance and Efficiency}
\begin{table*}[h!]
  \centering
  \caption{Performance on Benchmarks}
  \vskip -2ex
  \small
  \scalebox{0.75}{
  \begin{tabular}{|c|c|c|c|c|c|c|c|c|c|c|}
    \hline
    & F1/F1+ & BTS-2 & CraterLake & ARK & FAB & Poseidon & Over 100× & CL+MAD-32 &\cellcolor{gray!40}\textbf{FPGA-EFFACT}&\cellcolor{gray!40}\textbf{ASIC-EFFACT} \\
    \hline
    Parallelism & 2048 & 2048 & 2048 & 1024 & 256 & 256 & - & 2048 & 256 & 1024 \\
    \hline
    Multiplier Number & 18432 & 8192 & $\geq$33792 & 20480 & 256 & 256 & - & 14336 & \cellcolor{gray!40}\textbf{512} & \cellcolor{gray!40}\textbf{2048} \\
    \hline
    HBM Bandwidth & 1 TB/s & 1 TB/s & 1 TB/s & 1 TB/s & 460 GB/s & 460 GB/s & - & 1 TB/s & 460 GB/s& 1.2 TB/s \\
    \hline
    On-chip Memory Cap& 64 MB & 512 MB & 282 MB & 588 MB & 43 MB & 8.6 MB & - & 32 MB & \cellcolor{gray!40}\textbf{7.6 MB} & \cellcolor{gray!40}\textbf{27 MB} \\
    \hline
    Bootstrapping($T_{A.S.}$) & 260 us & 0.045 us & 0.017 us & 0.014 us & 0.477 us & 0.840 us & 0.74 us & ~0.270us & \cellcolor{gray!40}\textbf{0.566 us} & \cellcolor{gray!40}\textbf{0.0548 us} \\ 
    \hline
    HELR(1 iteration) & 1024 ms & 28.4 ms & 3.73 ms & 7.72 ms & 103 ms &  86.3 ms\footnotemark[1]  & 775 ms & 47.81 ms & \cellcolor{gray!40}\textbf{64.55 ms} & \cellcolor{gray!40}\textbf{8.7 ms} \\
    \hline 
    ResNet-20 & 2693 ms & 2020 ms & 249.45 ms & 294 ms & - &  2661.23 ms & - & {\color{rebuttal}1015.8} ms\footnotemark[2] & \cellcolor{gray!40}\textbf{2175.41 ms} &  \cellcolor{gray!40}\textbf{436.95 ms} \\
    \hline 
    DBLookup & 4.36 ms & - & - & - & - & -& - & - & \cellcolor{gray!40}\textbf{0.86 ms} & \cellcolor{gray!40}\textbf{0.13 ms}\\
    \hline
  \end{tabular}
    }

  \label{Benchmark}
\end{table*}
\vskip -2ex

{\color{rebuttal}We use the amortized time ($\textbf{T}_{A.S}$)\cite{jung_over_nodate} to evaluate the performance of Fully-packed Bootstrapping. The amortized time of Bootstrapping effectively captures the reciprocal throughput of a CKKS scheme with a certain parameter set. Therefore, it is widely used to evaluate the Bootstrapping performance in prior FHE accelerators\cite{kim_ark_2022,kim_bts_2022,agrawalheap}.}
ASIC-EFFACT is 13.49$\times$, 4743.79$\times$, 0.82$\times$, 0.31$\times$, 0.26$\times$, and 4.93$\times$ faster than GPU\cite{jung_over_nodate}, F1\cite{samardzic_f1_2021}, BTS~\cite{kim_bts_2022}, CraterLake~\cite{samardzic_craterlake_2022}, ARK~\cite{kim_ark_2022}, and MAD\cite{MAD} as shown in Table \ref{Benchmark} in CKKS. ASIC-EFFACT is slower than most ASIC designs in Bootstrapping since (1) ASIC-EFFACT works only at 500 MHz with fewer multipliers while others run at more than 1 GHz. (2) Bootstrapping features frequent data movement that blocks the parallelism. 
However, ASIC-EFFACT speeds up over MAD due to (1) our streaming optimization and global memory management reduce nearly 40\% DRAM access and the corresponding SRAM latency, (2) excessive static and dynamic scheduling well explores the instruction parallelism compared to MAD's hand-tuned data path, and (3) circuit-level reuse scheme is performed on the highly serial data path which results in acceleration with fewer computing resources.
While in HELR, ASIC-EFFACT is 89.1$\times$, 117.7$\times$, 3.26$\times$, 0.43$\times$, 0.89$\times$, and 5.5$\times$ faster than GPU, F1, BTS, CraterLake, ARK, and MAD. Unlike Bootstrapping, ASIC-EFFACT becomes more comparable with prior ASIC designs even if fewer function units and lower frequency are provided because HELR does not require as intermediate load/store as Bootstrapping as profiled by CraterLake\cite{samardzic_craterlake_2022}.
The same thing also happens in ResNet, in which ASIC-EFFACT is 6.16$\times$, 4.62$\times$, 0.57$\times$, 0.67$\times$, and 2.35$\times$ faster than F1, BTS, CraterLake, ARK, and MAD.
\begin{figure}
    \centering
    \includegraphics[width = 0.7\linewidth]{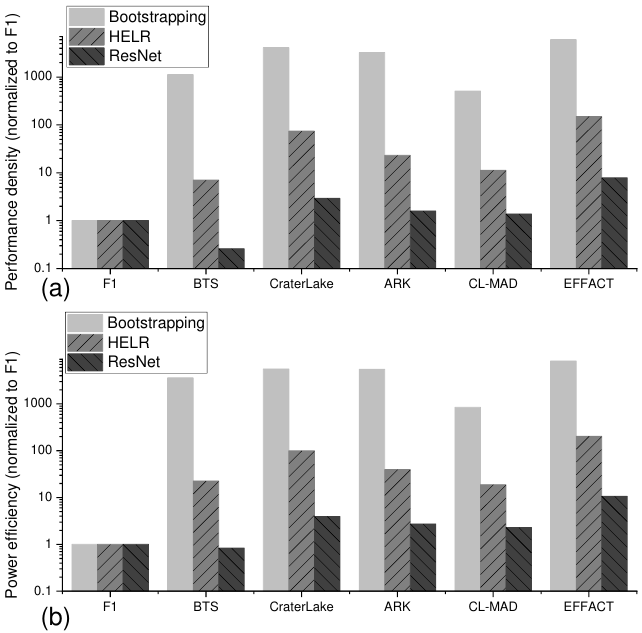}
     \vskip -2ex
    \caption{(a) Performance density comparison, (b) Power efficiency comparison.}
    \vskip -2ex
    \label{fig:density}
   
\end{figure}

\begin{figure}
    \centering
    \includegraphics[width = 0.75\linewidth]{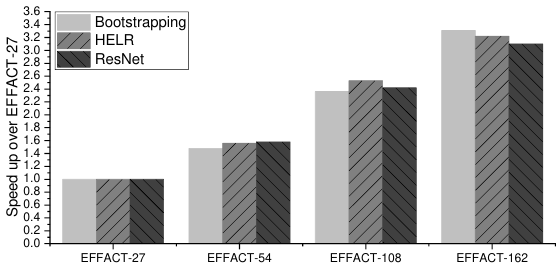}
     \vskip -2ex
    \caption{How performance scales up with memory and computing resources.}
    \label{fig:scale}
   
\end{figure}
Performance density is calculated by throughput per area to provide an area efficiency comparison and a scale-out evaluation. We scale prior ASIC designs to a 28-nm technology using the scaling technique provided by TSMC in\cite{14global,7sram,16sram}, in which we use our SRAM IP to evaluate the SRAM of CraterLake like SHARP\cite{sharp-kim} does since CraterLake has not reported the exact power of SRAM. Figure \ref{fig:density}(a) shows that in terms of performance density, ASIC-EFFACT achieves 6054.9$\times$, 5.35$\times$, 1.46$\times$, 1.86$\times$, and 11.89$\times$ higher performance density than F1, BTS, CraterLake, ARK, and MAD in Bootstrapping. While in the HELR and ResNet, ASIC-EFFACT outperforms prior designs at least by 2.02$\times$ and 2.7$\times$. 
We also look into the power efficiency evaluated by performance per Watt. Figure \ref{fig:density}(b) depicts that ASIC-EFFACT surpasses F1, BTS, CraterLake, ARK, and MAD by 8256.62$\times$, 2.28$\times$, 1.48$\times$, 1.49$\times$, and 9.76$\times$ respectively in Bootstrapping by scaling technology node. ASIC-EFFACT also outperforms prior architectures in HELR and ResNet by at least 2.04$\times$ and 2.72$\times$.   
ASIC-EFFACT achieves the best area and power efficiency due to (1) streaming access removing nearly 40\% DRAM fetches compared to MAD, thus, enhancing the SRAM efficiency, (2) computing resources re-assignment based on our analysis, and (3) efficient fine-grained function units and circuit-level reuse scheme further dramatically reducing the area without heavy performance penalty.
When applying SHARP's optimization\cite{sharp-kim} to EFFACT, we find that SAHRP-EFFACT will further improve EFFACT's area and power efficiency by 1.82$\times$ and 1.74$\times$ on average.

On the FPGA side, FPGA-EFFACT can well explore the parallelism between all the HE operations globally, thus, FPGA-EFFACT boosts both FAB\cite{agrawal_fab_2022} and Poseidon~\cite{yang_poseidon_nodate} in HELR by 1.59$\times$ and 1.34$\times$. We also observe that FPGA-EFFACT outperforms Poseidon in Bootstrapping by 1.48$\times$. However, 7.6-MB SRAM makes the frequent data movement during Bootstrapping worse for our compiler, thus, FPGA-EFFACT cannot exceed FAB in Bootstrapping.

\begin{figure}
    \centering
    \includegraphics[width = 0.8\linewidth]{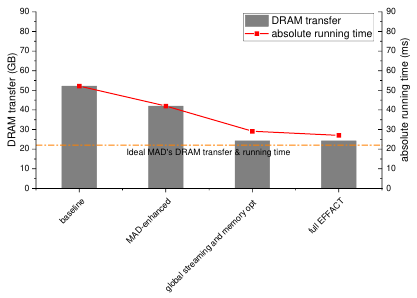}
     \vskip -2ex
    \caption{{\color{rebuttal}How Bootstrapping's DRAM transfer and absolute running time change when incrementally applying MAD's optimizations, EFFACT's global scheduling and streaming, and EFFACT's circuit reuse scheme.}}
    \label{fig:breakdown}
   
\end{figure}

\subsection{Scalibility of EFFACT}
To analyze how the computing resources and SRAM capacity impact EFFACT's performance, we create EFFACT-54 with 54-MB SRAM and 4096 multipliers, EFFACT-108 with 108-MB SRAM and 8192 multipliers, and EFFACT-162 with 162-MB SRAM and 12288 multipliers. The performance is evaluated using a cycle-accurate C++ simulator. Figure \ref{fig:scale} shows the results. When both the computing and SRAM resources scale up, EFFACT can directly have better performance since (1) when adding multipliers, the performance of NTT instructions increases linearly, (2) the serial normal MULT and ADD can be faster with the help of NTT units, and (3) larger SRAM incur fewer DRAM fetches, which means fewer memory stalls and more DRAM bandwidth can be provided to streaming optimization. According to the simulation results, EFFACT-108 can outperform ARK and CraterLake in HELR and ResNet. Since Bootstrapping is more memory intensive, EFFACT should have a 162-MB SRAM and 12288 multipliers to catch up with ARK and CraterLake.

\subsection{Application to other schemes}
Since the CKKS, BGV, and BFV share the same algebraic structure at the bottom layer, correspondingly, the same basic operations of the ciphertext (ModMult, ModAdd, NTT, Auto), it's obvious that EFFACT is also capable of accelerating the other two schemes, we evaluate ASIC-EFFACT and FPGA-EFFACT over BGV application, DBLookup. From Table \ref{Benchmark}, ASIC-EFFACT and FPGA-EFFACT are 33.54$\times$ and 5.07$\times$ faster than F1.
As for the boolean FHE scheme, take TFHE\cite{jain_threshold_nodate} as an example, EFFACT also demonstrates excellent acceleration capabilities. The key operation in TFHE is bootstrapping, which contains 3 major sub-operations, $Modulus\space Switching$, $Blind \space Rotation$, and $Sample \space Extraction$. {\color{rebuttal}$Modulus\space Switching$ can be mapped into modular arithmetic and NTT in EFFACT, even though FFT is used in previous work\cite{MATCHA}. For $Blind \space Rotation$ and $Sample \space Extraction$, as described in \cite{agrawalheap,chillotti_tfhe_2020}, when excluding the modular arithmetic and NTT, they are mainly linear shift operations with some slots being reversed. Therefore, EFFACT can support them using our automorphism unit by bypassing the fixed network and controlling the Muxes and reverse units to form shift operations and reverse data. We evaluate ASIC-EFFACT on TFHE Bootstrapping\cite{chillotti_tfhe_2020} under $N=2^{13}, logQ=218, h=1, l=2$ like HEAP\cite{agrawalheap}, and it shows the performance of 0.576-ms. Due to the flexibility of supporting different schemes, EFFACT has more potential than prior designs in the transciphering applications such as switching to AES\cite{e2e} or TFHE\cite{agrawalheap}.}

\footnotetext[1]{To complete the full HELR benchmark, Poseidon will need more bootstrapping operations rather than 10 iterations combined with 2 bootstrapping.} 
\footnotetext[2]{We evaluate the performance of CL+MAD-32 on ResNet-20 using SimFHE under the same parameter settings as EFFACT.}



{\color{rebuttal}\subsection{Sensitivity Study}

Since Bootstrapping operations represent a fundamental HE primitive, we analyze how our optimizations influence Bootstrapping's DRAM transfer and absolute running time to study how EFFACT enhances efficiency in a resource-constrained scenario. We create a bold baseline accelerator under our resource-constrained hardware settings and parameter set similar to ASIC-EFFACT (27-MB SRAM, 1TB/s DRAM bandwidth for simplification, 2048 modular multipliers, and 3072 modular adders) without any optimization and incrementally switch to MAD's and our optimizations. Figure \ref{fig:breakdown} shows the results.

We assume that the baseline, MAD-enhanced baseline, and ideal MAD have an ideal parallelism between memory operations and computations.  
The ideal MAD has unbounded computing resources and uses its infinite modular multipliers, modular adders and their buffers to support the smooth data flow processing and storing intermediate results, therefore no intermediate results are spilled and its main DRAM transfer is used to load secret keys or perform data structure transformation. However, MAD only looks into several data paths and highly relies on the number of computing resources and their buffers to consume the data flow results immediately. Therefore, with a restricted number of computing resources and their registers, MAD-enhanced baseline cannot perform as well as the ideal MAD. It only reduces the DRAM transfer and absolute running time by 1.24$\times$ compared to the non-optimized baseline.  
Our automatic scheduling and streaming optimization is applied at the whole program level and is aware of the limited computing resources and the small on-chip memory, therefore it not only reduces the 42.2\% DRAM transfer but also reduces 30.6\% of the absolute running time. When further applying our circuit-level NTT reuse scheme, it does not impact the DRAM transfer since it is only a computing optimization. Instead, it improves the normal ADD and MULT throughput which are shown in Figure \ref{op_anal}, leading to a 1.1$\times$ absolute running time improvement. 

}

\section{Related Work}
\textbf{CPU/GPU Acceleration. }Previous designs have taken a deep dive into the acceleration of HE primitives by making better use of the CPUs and GPUs. Many software libraries including Lattigo\cite{mouchet_lattigo_2020}, HElib\cite{halevi_design_nodate}, SEAL\cite{chen_simple_nodate}, HEAAN\cite{HEAAN}, and PALISADE\cite{noauthor_palisade_2022} have been proposed to improve the HE performance on CPUs. Intel HEXL\cite{intel-heax} further used the AVX-512 to accelerate HE operations over SEAL and PALISADE. However, due to the limited resources, CPU schemes remain impractical. Over 100$\times$\cite{jung_over_nodate} accelerated the FHE primitives including CKKS bootstrapping on GPUs by better utilizing the memory bandwidth.

\textbf{FPGA/ASIC Acceleration. }HEAX\cite{riazi_heax_2020} and \cite{sinha_roy_fpga-based_2019} proposed FPGA solutions for level HE. FAB\cite{agrawal_fab_2022} and Poseidon\cite{yang_poseidon_nodate} are pioneers in realizing the fully-packed bootstrapping on FPGA, {\color{gg_r}while HEAP\cite{agrawalheap} explored the possibility of FPGA acceleration for HE scheme-switching}. However, their designs either target leveled HE or never explore the parallelism between HE ops. CoFHEE\cite{nabeel_cofhee_2022}, F1\cite{samardzic_f1_2021}, BTS\cite{kim_bts_2022}, ARK\cite{kim_ark_2022}, Cheetah\cite{reagen_cheetah_2021}, {\color{gg_r}SHARP\cite{sharp-kim}, }and CraterLake\cite{samardzic_craterlake_2022} are ASIC designs to accelerate HE schemes. Cheetah targets privacy-preserving ML using LHE and requires expensive communication overhead. F1, BTS, ARK, SHARP, and CraterLake support fully-packed bootstrapping and show impressive acceleration over GPUs and FPGAs. However, they require a huge amount of on-chip SRAMs and computing resources. {\color{rebuttal} CiFlow\cite{neda2024ciflowdataflowanalysisoptimization} proposes a novel data flow framework tuned for the hybrid key switch and shows huge data movement reduction. However, their data flow is manually tuned only for the evaluation key's DRAM loading in the key-switching. It highly relies on the large number of computing resources to immediately consume their intermediate results of their data flow.}

\section{Conclusion}
{\color{gg}
In this work, we propose EFFACT, aiming at the challenging problems of highly efficient and cost-sensitive FHE acceleration and propose.
EFFACT analyzes the proportion of different FHE operations and the inherent parallelism within them in several real-world benchmarks, exhibiting opportunities to re-assign the computing resources. 
Based on this observation, we tailor the specialized BConv unit, devise novel compact NTT and automorphism units, and propose a circuit-level reuse scheme. These innovations aim to leverage reconfigurable technology to enhance efficiency while minimizing performance penalties.
On-chip memory is also significant for a compact accelerator.
To fully utilize the limited SRAM, we propose a streaming optimization in which function units can directly fetch their operands from DRAM instead of waiting for SRAM to be filled.
Also, we perform a design space exploration to find an optimal SRAM size targeting high efficiency and low cost.
Without loss of generality, we also devise a generalized ISA and a compiler backend that can support several FHE schemes and can be integrated into recent compiler frontends.
Experimental results demonstrate that EFFACT outperforms state-of-the-art baseline FHE accelerators in terms of efficiency, and area/on-chip memory overhead. 
}


\section*{Acknowledgements}

This work is supported in part by the National Key R\&D Program of China (Grant No. 2023YFB4403500), and in part by the National Natural Science Foundation of China (Grant No. 62274102)

\bibliographystyle{IEEEtranS}
\bibliography{references}

\end{document}